\documentclass[useAMS]{emulateapj}

\pdfoutput=1
\usepackage{epsfig}
\usepackage{amsmath}
\usepackage{amssymb}
\usepackage{appendix}

\def\be{\begin{equation}}
\def\ee{\end{equation}}
\def\ba{\begin{eqnarray}}
\def\ea{\end{eqnarray}}
\def\go{\mathrel{\raise.3ex\hbox{$>$}\mkern-14mu
             \lower0.6ex\hbox{$\sim$}}}
\def\lo{\mathrel{\raise.3ex\hbox{$<$}\mkern-14mu
             \lower0.6ex\hbox{$\sim$}}}

\def\cE{{\cal E}}
\def\cR{{\cal R}}
\def\cI{{\cal I}}
\def\bmu{{\mbox{\boldmath $\mu$}}}
\def\bphi{{\mbox{\boldmath $\phi$}}}

\begin{document}

\title{DC Circuit Powered by Orbital Motion: Magnetic Interactions in 
Compact Object Binaries and Exoplanetary Systems}

\author{Dong Lai}
\affil{Department of Astronomy, Cornell University,
    Ithaca, NY 14850}

\begin{abstract}
The unipolar induction DC circuit model, originally developed by
Goldreich \& Lynden-Bell for the Jupiter-Io system, has been applied
to different types of binary systems in recent years.
We show that there exists an upper limit to the magnetic interaction
torque and energy dissipation rate in such model. This arises
because when the resistance of the circuit is too small, 
the large current flow severely twists
the magnetic flux tube connecting the two binary components, leading
to breakdown of the circuit. Applying this limit,
we find that in coalescing neutron star 
binaries, magnetic interactions produce negligible correction to 
the phase evolution of the gravitational waveform, even 
for magnetar-like field strengths. However, energy dissipation 
in the binary magnetosphere may still give rise to electromagnetic radiation 
prior to the final merger.
For ultra-compact white dwarf binaries, we find that DC circuit does not 
provide adequate energy dissipation to explain the observed X-ray luminosities
of several sources. For exoplanetary systems containing close-in Jupiters or 
super-Earths, magnetic torque and dissipation are negligible, except possibly
during the early T Tauri phase, when the stellar magnetic field is
stronger than $10^3\,$G.
\end{abstract}

\section{Introduction}

In a seminal paper, Goldreich \& Lynden-Bell (1969) developed a DC
circuit model for the magnetic interaction between Jupiter and its
satellite Io (see also Piddington \& Drake 1968).  In this model, the
orbital motion of Io in the rotating magnetosphere of Jupiter
generates electromotive force (EMF), driving a DC current between Io
and Jupiter, with the closed magnetic field lines serving as wires. In
essence, the Jupiter-Io system operates as a unipolar inductor. This
model helped to explain Jupiter's decametric radio emissions which are
correlated with the motion of Io and the hot spot in Jupiter's
atmosphere that is linked magnetically to Io (e.g., Clarke et
al.~1996).

In recent years, the DC circuit model of Goldreich \& Lynden-Bell
has been adapted and applied to other types of astrophysical
binary systems, including (i) Planets around magnetic white dwarfs 
(Li et al.~1998); (ii) Ultra-compact white dwarf binaries with periods
ranging from a few minutes to a hour (Wu et al. 2002; Dall'Osso et
al.~2006,2007); (iii) Coalescing neutron star - black Hole binaries
and neutron star - neutron star binaries prior to the final merger
(McWilliams \& Levin 2011; Piro 2012; see also Vietri 1996, Hansen \&
Lyutikov 2001 and Lyutikov 2011 for more general discussion of
unipolar induction in such binaries); (iv) Exoplanetary systems
containing short-period ($\sim$ days) planets orbiting magnetic MS stars
(Laine \& Lin 2012);

In the DC circuit model, the magnetic interaction torque and the
related energy dissipation rate are inversely proportional to the
total resistance $\cR_{\rm tot}$ of the circuit, including
contributions from the two binary components and the
magnetosphere. Thus it seems that extreme power may be produced for
highly conducting binary systems.  However, we show in this paper that
when $\cR_{\rm tot}$ is smaller than a critical value ($\sim 480v_{\rm
  rel}/c$~Ohms, where $v_{\rm rel}$ is the orbital velocity as seen in
the rotating frame of the magnetic star), the large current flowing in
the circuit will break the magnetic connection between the binary
components. Thus there exists an upper limit to the magnetic torque
and the energy dissipation rate. In Sects.~2-4, we examine the general
aspect of the DC circuit model, the upper limit and its connection
with ``open circuit'' effect (``Alfven radiation'').  We then discuss
applications of the model to various astrophysical binaries in Sects.~5-8
and conclude in Sect.~9.

\section{DC Circuit Model of Magnetic Interaction}

Consider a binary system consisting of a magnetic star (the
``primary'', with mass $M_\star$, radius $R_\star$, spin $\Omega_s$, and magnetic
dipole moment $\mu$) and a non-magnetic companion (mass $M_c$, radius
$R_c$). The orbital separation is $a$, and the orbital angular
frequency is $\Omega$.  The magnetic field strength at the surface of
the primary is $B_\star=\mu/R_\star^3$.  The whole binary system is embedded
in a tenuous plasma (magnetosphere).  For simplicity, we assume
${\bf\Omega}$, ${\bf\Omega_s}$ and $\bmu$ are all aligned.

\begin{figure}
\includegraphics[scale=0.31]{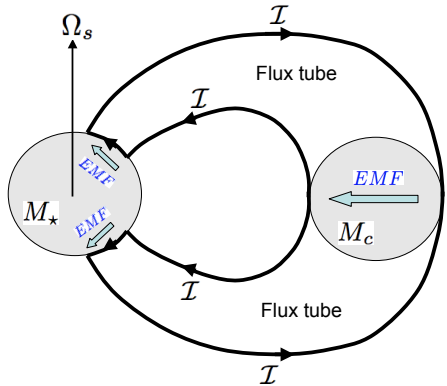}
\vskip -0.7cm
\caption{\label{fig1} DC circuit model of magnetic interactions in binary systems
{\it a la} Goldreich \& Lynden-Bell (1969).}
\end{figure}

The motion of the non-magnetic companion relative to the magnetic field of the
primary produces an EMF $\cE \simeq 2R_c |E|$, where ${\bf E}=
{\bf v}_{\rm rel}\times {\bf B}/c$, with 
${\bf v}_{\rm rel}=(\Omega-\Omega_s)a\,{\hat\bphi}$ and 
${\bf B}=(-\mu/a^3){\hat{\bf z}}$.
This gives
\be
\cE\simeq {2\mu R_c\over ca^2}\Delta\Omega,
\label{eq:emf}\ee
where $\Delta\Omega=\Omega-\Omega_s$. The EMF drives a current along the magnetic
field lines in the magnetosphere, connecting the primary and the companion
through two flux tubes (Fig.~1). The current in the circuit is 
\be
\cI={\cE\over \cR_{\rm tot}},
\ee
where the total resistance of the circuit is
\be
\cR_{\rm tot}=\cR_\star+\cR_c+2\cR_{\rm mag},
\ee
with $\cR_\star,\,\cR_c,\,\cR_{\rm mag}$ the resistances of the magnetic
star, the companion and the magnetosphere, respectively.
The energy dissipation rate of the system is then
\be
\dot E_{\rm diss}=2{\cal I}^2R_{\rm tot}={2\cE^2\over\cR_{\rm tot}},
\label{eq:diss}\ee
where the factor of 2 accounts for both the upper and lower sides of the circuit.

Note that because of the finite resistances $\cR_\star$ and $\cR_c$, the magnetic
flux tube in general rotates at a rate $\Omega_F$ between $\Omega_s$ and $\Omega$.
The slippage of the companion relative to the magnetic field gives $\cE_c\simeq
(2\mu R_c/ca^2)(\Omega-\Omega_F)$. The slippage of the primary's rotation relate to the 
flux tube generates $\cE_\star\simeq (2\mu R_c/ca^2)(\Omega_F-\Omega_s)$ at the magnetic
foot on the surface of the primary. The total EMF $(\cE_\star+\cE_c)$
is the same as in Eq.~(\ref{eq:emf}), independent of $\Omega_F$.

The total magnetic force (in the azimuthal direction) on the companion is 
$F_\phi\simeq (2R_c) (2{\cal I}B_z/c)$, with $B_z=-\mu/a^3$. 
Thus the torque acting on the binary's orbital angular momentum is
\be
T=\dot J_{\rm orb}\simeq {4\over c}a\,R_c{\cal I}B_z\simeq
-{4\mu R_c\over ca^2}{\cE\over {\cal R}_{\rm tot}}.
\ee
The torque on the primary is $I_\star\dot\Omega_s=-T$ (where $I_\star$ is
the moment of inertia).
The orbital energy loss rate associated with $T$ is then
\be
\dot E_{\rm orb}=T\Omega\simeq
-{2\cE^2\over {\cal R}_{\rm tot}}\,{\Omega\over\Delta\Omega}.
\ee
The total energy dissipation rate of the system is
\be
\dot E_{\rm diss}=\left(-\dot E_{\rm orb}\right)-
I_\star\Omega_s\dot\Omega_s
=-T\Delta\Omega={2\cE^2\over \cR_{\rm tot}},
\ee
in agreement with Eq.~(\ref{eq:diss}).

\section{Maximum Torque and Dissipation}

The equations in the previous section clearly show that the binary
interaction torque and energy dissipation associated with the DC 
circuit increase with decreasing total resistance $\cR_{\rm tot}$.
Is there a problem for the DC model when $\cR_{\rm tot}$ is too small?
The answer is yes.

The current in the circuit produces toroidal magnetic field, which
has the same magnitude but opposite direction above and below 
the equatorial plane. The toroidal field just above the 
companion star (in the upper flux tube) is $B_{\phi+}\simeq -(2\pi/c){\cal J}_r$,
where the (height-integrated) surface current in the companion is
${\cal J}_r\sim -4\cI/(\pi R_c)$. Thus the azimuthal twist of the
flux tube is
\be
\zeta_\phi=-{B_{\phi+}\over B_z}={8\cE\over cR_c\cR_{\rm tot}|B_z|}
={16 v_{\rm rel}\over c^2\cR_{\rm tot}},
\label{eq:zeta}\ee
where $v_{\rm rel}=a\Delta\Omega=a(\Omega-\Omega_s)$ is the orbital
velocity in the corotating frame of the primary star.
Clearly, when $\cR_{\rm tot}$ is less than $16v_{\rm rel}/c^2$,
the flux tube will be highly twisted.

GL already speculated in 1969 that the DC circuit would break
down when the twist is too large. (For the Jupiter-Io system
parameters adopted by GL, the twist $|\zeta_\phi|\ll 1$.) Since then,
numerous works have confirmed that this is indeed the case.
Theoretical studies and numerical simulations, usually carried out 
in the contexts of solar flares
and accretion disks, have shown that as a flux tube is twisted beyond
$\zeta_\phi\go 1$, the magnetic pressure associated with $B_\phi$
makes the flux tube expand outward and the magnetic fields open up,
allowing the system to reach a lower energy state (e.g., Aly 1985; Aly
\& Kuijpers 1990; van Ballegooijen 1994; Lynden-Bell \& Boily 1994;
Lovelace et al.~1995; Uzdensky et al.~2002).
Thus, a DC circuit with $\zeta_\phi\go 1$ cannot be realized: The
flux tube will break up, disconnecting the linkage between the two 
binary components.

A binary system with $\cR_{\rm tot}\lo 16v_{\rm rel}/c^2$ 
cannot establish a steady-state DC circuit. 
The electrodynamics is likely rather complex.
At best we can expect a quasi-cyclic circuit, involving several steps:
(a) The magnetic field from the primary penetrates
part of the companion, establishing magnetic linkage between the two
stars; (b) The linked fields are twisted by differential rotation, generating
toroidal field from the linked poloidal field; (c) As the toroidal magnetic field
becomes comparable to the poloidal field, the fields inflate and 
the flux tube breaks, disrupting the magnetic linkage; 
(d) Reconnection between the inflated field lines relaxes the shear and restore
the linkage. The whole cycle repeats.

In any case, we can use the dimensionless azimuthal twist $\zeta_\phi$
to parameterize the magnetic torque and energy dissipation rate:
\ba
&& T= -\zeta_\phi{\mu^2 R_c^2\over 2a^5},\label{eq:tmax}\\
&& \dot E_{\rm diss} = -T \Delta\Omega 
=\zeta_\phi\Delta\Omega {\mu^2 R_c^2\over 2a^5}.
\label{eq:emax}\ea
The maximum torque and dissipation are obtained by setting $\zeta_\phi\sim 1$.
Note that in the above, $T$ is negative since we
are assuming $\Omega>\Omega_s$. A reasonable extension would
let $\zeta_\phi=\zeta (\Delta\Omega)/\Omega$, with $\zeta>0$.

\section{Alfven Drag}

As discussed in GL, the validity of the DC circuit model requires that the
slippage of the flux tube relative to the companion during the
round-trip Alfven travel time ($t_A$) along the flux tube be much less
than $R_c$, i.e., $(\Omega-\Omega_F)a t_A\ll R_c$. When this condition
is not satisfied or when the poloidal field opens up, 
the disturbance generated by the companion's orbital
motion ${\bf v}_{\rm rel}$ will propagate along the field line as
Alfven waves and radiate away (Drell, Foley \& Ruderman 1965). The
Alfven radiation power associated with the ``open circuit'' is
\be
\dot E_{\rm Alf}\sim {1\over 4\pi}\left(B_z v_{\rm rel}/v_A\right)^2v_A (2\pi R_c^2)
={1\over 2}B_z^2R_c^2\,{v_{\rm rel}^2\over v_A},
\ee
where 
$v_A$ is the Alfven speed in the magnetosphere
(assuming $v_{\rm rel}\lo v_A$). The associated drag force is simply 
$F_\phi=-\dot E_{\rm Alf}/v_{\rm rel}$, and the torque on the orbit is
\be
T_{\rm Alf}\sim 
-{1\over 2}B_z^2R_c^2a\,{v_{\rm rel}\over v_A}.
\ee
Comparing $\dot E_{\rm Alf}$ with Eq.~(\ref{eq:emax}), we find
\be
{\dot E_{\rm Alf}\over \dot E_{\rm diss}}\sim {v_{\rm rel}\over\zeta_\phi v_A}.
\label{eq:ealf}\ee
Thus, $\dot E_{\rm Alf}$ is always smaller than the maximum $\dot
E_{\rm diss}$ of a DC circuit. Equation~(\ref{eq:emax}) (with
$\zeta_\phi\sim 1$) represents the maximum magnetic dissipation rate
of the binary system, regardless of the details of the
electrodynamics.

\section{Neutron Star - Neutron Star Binaries}

Gravitational wave (GW) emission drives the orbital decay of the neutron star
(NS) binary, with timescale
\be
t_{\rm GW}={a\over |\dot a|}={5c^5a^4\over 64G^3M_\star^3 q(1+q)}
\simeq 0.012\left(\!{a\over 30\,{\rm km}}\!\right)^{\!4}{\rm s},
\label{eq:tgw}\ee
where in last equality we have adopted $M_\star=1.4M_\odot$ and mass ratio 
$q=M_c/M_\star=1$.
The magnetic torque tends to spin up the primary when $\Omega>\Omega_s$.
Spin-orbit synchronization is possible only if 
the synchronization time $t_{\rm syn}=I_\star\Omega/|T|$ is less than
$t_{\rm GW}$ at some orbital radii.  With 
$I_\star=\kappa M_\star R_\star^2$ and the torque (\ref{eq:tmax}), 
we find
\ba
t_{\rm syn}&=&{2\kappa(1+q)\over\zeta_\phi\Omega}
\left(\!{GM_\star^2\over B_\star^2R_\star^4}\!\right)
\!\left(\!{a\over R_c}\!\right)^2\nonumber\\
&\simeq&2\times 10^7\zeta_\phi^{-1}\!\left(\!{B_\star\over 10^{13}\,{\rm G}}
\!\right)^{\!-2}\!\left({a\over 30\,{\rm km}}\right)^{7/2}{\rm s},
\label{eq:tsyn}\ea
where in the second line we have adopted $\kappa=0.4$ and $R_\star=R_c=10$~km.
Clearly, even with magnetar-like field strength ($B_\star\sim 10^{15}$~G) and
maximum efficiency ($\zeta_\phi\sim 1$), spin-orbit synchronization cannot be
achieved by magnetic torque. [It was already known that tidal torque, due to both
equilibrium tide (Bildsten \& Cutler 1992) and resonant tides (Lai 1994), cannot
synchronize the NS spin during binary inspiral.]

For the same reason, the effect of magnetic torque on the 
number of GW cycles during binary inspiral, $N$, is small. We find
\be
{dN\over d\ln f}={1\over 1+\alpha}\left({dN\over d\ln f}\right)_0,
\ee
where $f=\Omega/\pi$ is the GW frequency, and 
\be
\left(dN\over d\ln f\right)_0={5c^5(1+q)^{1/3}\over 96\pi q M_\star^{5/3}
(\pi Gf)^{5/3}}
\ee
is the usual leading-order point-mass GW cycles.
The correction factor due to the magnetic torque is
\be
\alpha=
{2t_{\rm GW}\over J_{\rm orb}/|T|}=
{2I_\star\over \mu_m a^2}\left({t_{\rm GW}\over t_{\rm syn}}\right),
\ee
where $\mu_m=M_\star q/(1+q)$ is the reduced mass of the binary.
With Eqs.~(\ref{eq:tgw})-(\ref{eq:tsyn}), we see that GW phase error
$\alpha(dN/d\ln f)_0$ is much less than unity even for 
$B_\star\sim 10^{15}$~G and maximum $\zeta_\phi\sim 1$.


The energy dissipation rate is 
\ba
&& \dot E_{\rm diss}=\zeta_\phi\left(\!{v_{\rm rel}\over c}\!\right){B_\star^2R_\star^6
R_c^2c\over 2a^6}\nonumber\\
&&\quad = 7.4\times 10^{44}\zeta_\phi\left(\!{B_\star\over 
10^{13}\,{\rm G}}\!\right)^{\!2}\!\left(\!{a\over 30\,{\rm km}}\!\right)^{\!\!-13/2}
\!{\rm erg\,s}^{-1},
\ea
where in the second line we have used $v_{\rm rel}\simeq r\Omega$ (for 
$\Omega_s\ll \Omega$) and adopted canonical parameters
($M_\star=M_c=1.4M_\odot$, $R_\star=R_c=10$~km).
The total energy dissipation per $\ln a$ is 
\ba
{dE_{\rm diss}\over d\ln a}&=&\dot E_{\rm diss}t_{\rm GW}\nonumber\\
&\simeq & 8.9\times 10^{42}\zeta_\phi\!\left(\!{B_\star\over 
10^{13}\,{\rm G}}\!\right)^{\!2}\!\left({a\over 30\,{\rm km}}\right)^{\!\!-5/2}
\!{\rm erg}.
\ea
Some fraction of this dissipation will emerge as electromagnetic
radiation counterpart of binary inspiral. Whether it is detectable at
extragalactic distance depends on the microphysics in the
magnetosphere, including particle acceleration and radiation mechanism
(e.g., Vietri 1996; Hansen \& Lyutikov 2001).

Piro (2012) recently applied the DC circuit model to NS binaries.
In the case when the magnetic NS primary dominates the total resistance, he
found that the magnetic torque synchronizes the NS spin prior to merger
and significantly affects the gravitational wave cycles, even for modest
($\sim 10^{12}$~G) NS magnetic fields. Using the resistance computed by 
Piro, we find from Eq.~(\ref{eq:zeta}) that the corresponding 
azimuthal twist $\zeta_\phi$ is much larger (by a factor $\sim 10^8$
at $\Omega=100$~s$^{-1}$) than unity, violating the upper limit discussed
in Sect.~3. 

If one assumes that the magnetosphere resistance is given by the
impedance of free space, $\cR_{\rm mag}=4\pi/c$, then the corresponding twist
is $\zeta_\phi=2v_{\rm rel}/(\pi c)$, which satisfies our upper limit.
The energy dissipation rate is then
\ba
&& \dot E_{\rm diss}=\left(\!{v_{\rm rel}\over c}\!\right)^2\!
{B_\star^2R_\star^6 R_c^2c\over \pi a^6}\nonumber\\
&&\quad = 1.7\times 10^{44}\left(\!{B_\star\over 10^{13}\,{\rm G}}\!\right)^{\!2}
\!\left({a\over 30\,{\rm km}}\right)^{\!\!-7}{\rm erg/s}.
\ea
Not surprisingly, this is approximately the same as the Alfven power 
$\dot E_{\rm Alf}$ [see Eq.~(\ref{eq:ealf})] with $v_A=c$ and
is in agreement with the estimate of Lyutikov (2011).

\section{Neutron Star - Black Hole Binaries}

The situation is similar to the case of NS-NS binaries.
In the membrane paradigm (Thorne et al.~1986), a black hole (BH)
of mass $M_H$ resembles a sphere of radius $R_c=R_H=2GM_H/c^2$ 
(neglecting BH spin)
and impedance $\cR_H=4\pi/c$. Neglecting the resistances of the
magnetosphere and the NS, the azimuthal twist of the flux tube in the DC 
circuit is 
\be
\zeta_\phi={4v_{\rm rel}\over \pi c},
\ee
which satisfies our upper limit (Sect.~3).
The energy dissipation rate is (cf. Lyutikov 2011; McWilliams \& Levin 2011)
\ba
&& \dot E_{\rm diss}=\left(\!{v_{\rm rel}\over c}\!\right)^2\!
{2B_\star^2R_\star^6 R_H^2c\over \pi a^6}\nonumber\\
&&\simeq 5.7\!\times\! 10^{42}\!\left(\!{B_\star\over 10^{13}\,{\rm G}}\!\right)^{\!2}
\!\!\left(\!{M_H\over 10M_\odot}\!\right)^{\!\!\!-4}
\!\!\!\left(\!{a\over 3R_H}\!\right)^{\!\!-7}\!\!\!{\rm erg\,s}^{-1},
\ea
where we have assumed $M_{BH}/M_\star\gg 1$.

\section{Ultra-compact Double White Dwarf Binaries}

Wu et al.~(2002) and Dall'Osso et al.~(2006,2007) (see also Wu 2009)
developed the DC circuit model for ultra-compact double white dwarf
(WD) binaries, particularly for the systems RX J1914+24 (period
$P=9.5$~mins) and RX J0806+15 ($P=5.4$~mins). Usual mass transfer
models appear to have difficulties explaining some of the properties
of these systems (e.g., the observed orbital decay). The DC circuit
model seeks to account for the observed X-ray luminosity
($10^{35}-10^{36}$~erg~s$^{-1}$ for RX J1914+24 assuming a distance of
100~pc) without mass accretion, while allowing for orbital decay
driven by gravitational radiation.

Using Eq.~(\ref{eq:emax}) with parameters appropriate to compact WD binaries,
we find 
\ba
&&\dot E_{\rm diss}=3.8\times 10^{29}\zeta_\phi \left(\!{\Delta\Omega\over\Omega}\!
\right)\left(\!{\mu\over 10^{32}\,{\rm G\,cm^3}}\!\right)^2
\left(\!{R_c\over 10^4\,{\rm km}}\!\right)^{\!2}\nonumber\\
&&\qquad \times 
\left(\!{M_\star+M_c\over 1\,M_\odot}\!\right)^{\!\!-5/3}
\left(\!{P\over 10\,{\rm min}}\!\right)^{\!\!-13/3}\!{\rm erg~s}^{-1}.
\ea
Note that $\mu=10^{32}$~G\,cm$^3$ corresponds to $B_\star\simeq 0.5$~MG (for 
$R_\star=6000$~km), approaching the field strengths of Intermediate Polars.
Obviously, even with the maximum asynchronization ($\Delta\Omega/\Omega=1$)
and maximum efficiency ($\zeta_\phi\sim 1$), $\dot E_{\rm diss}$ falls
far short of the observed X-ray luminosities. Wu et al.~(2002) calculated
the resistance of the WD and used Eq.~(\ref{eq:diss}) to obtain a much higher
energy dissipation power (see their Fig.~3) -- evidently, their result
(which was also adopted by Dall'Osso et al.~2006) corresponds to $\zeta_\phi\gg 1$,
violating our upper limit.

\section{Close-in Exoplanetary Systems}

Laine \& Lin (2012) recently applied the DC circuit model to study the
interaction of close-in planets with the magnetosphere of host
stars. From their calculation of the resistances of the planet, host star
and magnetosphere, they suggested that magnetic interaction may affect 
the orbital evolution of close-in super-Earths on a few Myr timescale, and 
produce hot spots on the surface of the host stars.

Applying Eq.~(\ref{eq:tmax}) to a planetary system ($M_c=M_p$, $R_c=R_p$), we
find that the magnetic torque induces orbital decay (assuming $\Omega>\Omega_s$,
as is the case for close-in planets in a few day orbit) at the rate 
\be
{\dot a\over a}=-{\zeta_\phi\mu^2 R_p^2\over a^5M_p(GM_\star a)^{\!1/2}}.
\ee
The timescale is 
\ba
&&{a\over |\dot a|}=5.7\times 10^{15}\zeta_\phi^{-1}
\left(\!{M_{\star}\over 1\,M_\odot}\!\right)^{\!1/2}
\!\left({R_{\star}\over 1\,R_\odot}\!\right)^{\!-6}
\!\left(\!{B_\star\over 1\,{\rm G}}\!\right)^{\!-2}\nonumber\\
&&\qquad \times \left(\!{R_p\over 1\,R_J}\!\right)\!
\left(\!{\bar\rho_p\over 1\,{\rm g\,cm}^{-3}}\!\right)\!
\left({a\over 0.04\,{\rm AU}}\right)^{\!11/2}{\rm yrs},
\ea
where $\bar\rho_p$ is the mean density of the planet. Thus it is clear
that even at maximum efficiency ($\zeta_\phi\sim 1$), a hot
Jupiter ($R_p\sim R_J$) or super-Earth ($R_p\sim 0.1R_J$) in a $P\sim
3$~d orbit ($a\simeq 0.04$~AU) around a solar-type star (with a
typical dipole field $B_\star$ of a few Gauss) will suffer negligible
orbital decay due to the magnetic torque. If super-Earths
migrate to their close-in locations in the early, T-Tauri phase of the star, 
when the stellar magnetic field is stronger (a few kG), they may experience 
appreciable orbital evolution due to magnetic interactions.

The energy dissipation rate is [Eq.~(\ref{eq:emax})]
\ba
&&\dot E_{\rm diss}=9.5\times 10^{20}\zeta_\phi 
\left(\!{\Delta\Omega\over\Omega}\!\right)\!
\left(\!{M_\star\over 1\,M_\odot}\!\right)^{\!\!1/2}\!\!
\left(\!{R_\star\over 1\,R_\odot}\!\right)^{\!6}\!
\left(\!{B_\star\over 1\,{\rm G}}\!\right)^2\nonumber\\
&&\qquad\times
\left(\!{R_p\over 1\,R_J}\!\right)^{\!2}
\left(\!{a\over 0.04\,{\rm AU}}\!\right)^{\!\!-13/2}\!{\rm erg~s}^{-1}.
\ea
If this energy is accumulated (e.g., building up twist from $\zeta_\phi\sim 0$ to
$\zeta_\phi\sim 1$) over time $\Delta t\simeq 2\pi/\Delta\Omega\sim 3$~d and then 
released suddenly ($\sim$ hours), the energy release is $\sim \dot E_{\rm diss}
\Delta t/2\sim 10^{26}$~erg (for the canonical parameter values adopted in the above
equation). This is much smaller than the energy release of solar flares
($10^{29}-10^{32}$~ergs) and of superflares from solar-type stars
($\go 10^{33}$~ergs; see Maehara et al.~2012).

Finally, it is instructive to compare the magnetic torque $T_{\rm mag}=T$
[Eq.~(\ref{eq:tmax})] with the tidal torque (due to tide raised
on the star by the planet). Parameterizing tidal dissipation by the quality factor 
$Q_\star$, we have 
\be
|T_{\rm tide}|=\left({9\over 4Q_\star'}\right){GM_p^2R_\star^5\over a^6},
\ee
where $Q_\star'=3Q_\star/(2k_2)$ and $k_2$ is the Love number (Goldreich \& Soter 1966).
Thus,
\ba
&&{|T_{\rm mag}|\over |T_{\rm tide}|}=3\times 10^{-12}\zeta_\phi Q_\star'
\left(\!{R_\star\over 1\,R_\odot}\!\right)\!
\left(\!{B_\star\over 1\,{\rm G}}\!\right)^{\!2}\nonumber\\
&&\qquad\times
\left(\!{R_p\over 1\,R_J}\!\right)^{\!\!-4}\!
\left(\!{\bar\rho_p\over 1\,{\rm g\,cm}^{-3}}\!\right)^{\!-2}\!
\left({a\over 0.04\,{\rm AU}}\right).
\ea
With $Q_\star'\sim 10^6-10^8$, the magnetic torque generally 
cannot compete with the tidal torque.

\section{Conclusion}

Closed DC circuit connecting a magnetic primary and a nonmagnetic
companion (Sect.~2) can be an efficient unipolar engine in a binary
system, potentially more efficient than an ``open'' circuit engine
(see Sect.~4). The power of this DC engine is inversely proportional
to the total resistance $\cR_{\rm tot}$ of the circuit.  However, we
have shown that when $\cR_{\rm tot}$ is less than a critical value,
$\sim 16v_{\rm rel}/c^2$ (in the cgs units) [see Eq.~(\ref{eq:zeta})]
or $480(v_{\rm rel}/c)$~Ohms, the magnetic flux tube connecting the two
binary components will be highly distorted and the circuit will
break. In this case, a quasi-cyclic unipolar engine may operate in the
system (Sect.~3). Thus, there exists an upper limit to the magnetic
interaction torque and the associated energy dissipation rate
[Eqs.~(\ref{eq:tmax})-(\ref{eq:emax})] of any unipolar engine in
magnetic binaries.  Several previous applications of the DC circuit
model to different types of binary systems apparently violated this
upper limit.  We have shown that:

(i) In coalescing double neutron star (NS) or NS - black hole
binaries: Magnetic interactions cannot synchronize the NS spin with
the orbital motion and have a negligible effect on phase evolution of
the gravitational waveform, even for magnetar field strengths.
Nevertheless, the energy dissipation associated with the unipolar engine
can produce electromagnetic radiation prior to binary merger, although the
detectability of this radiation depends on the microphysics of the 
binary magnetosphere.

(ii) In Ultra-compact white dwarf binaries:
Magnetic energy dissipation is too small to account for
the observed X-ray luminosities. Thus, the puzzling behaviors
of several sources (RX J1914+24 and RX J0806+15) cannot be
explained by the unipolar inductor circuit model.

(iii) In close-in exoplanetary systems: Interaction between 
hot Jupiters or Super-Earths and the magnetosphere of
their host stars does not lead to appreciable orbital evolution,
with the possible exception of the early T Tauri phase, when
the stellar dipole magnetic field is higher than $10^3$~Gauss.
Magnetic energy dissipation induced by the orbital motion of planets
is generally negligible compared to the
observed energy releases in stellar flares or superflares.


{\it Acknowledgments:}
I thank Doug N.C. Lin for many valuable discussions on this topic during
the the winter of 2010-2011.  This work has been supported in part by
the grants NSF AST-1008245, NASA NNX12AF85G and NNX10AP19G.

\def\apj{{Astrophys. J.}}
\def\apjs{{Astrophys. J. Supp.}}
\def\mnras{{Mon. Not. R. Astr. Soc.}}
\def\prl{{Phys. Rev. Lett.}}
\def\prd{{Phys. Rev. D}}
\def\apjl{{Astrophys. J. Let.}}
\def\pasp{{Publ. Astr. Soc. Pacific}}
\def\aa{{Astr. Astr.}}
\def\aapr{{Astr. Astr. Rev.}}


\end{document}